\documentclass[prd,preprint,superscriptaddress,floatfix,eqsecnum,nofootinbib,12pt]{revtex4}
\pdfoutput=1

\usepackage[english]{babel}
\usepackage{amsmath}
\usepackage{amssymb}
\usepackage{amsfonts}
\usepackage{color}
\usepackage{graphicx}
\usepackage{hyperref}
\usepackage{enumitem}

\def\be{\begin{equation}}
\def\ee{\end{equation}}
\def\beq{\begin{eqnarray}}
\def\eeq{\end{eqnarray}}
\def\p0{\phi_0}
\def\z0{\zeta_0}
\def\vx{{\vec x}}

\def\lpm{\lambda_{\pm}}
\def\th{{\tilde h}}
\def\epa{{(0)}}
\def\epb{{(2)}}

\begin{document}

\title{Primordial fluctuations from complex AdS saddle points}
\author{Thomas Hertog\footnote{thomas.hertog@fys.kuleuven.be} and Ellen van der Woerd\footnote{ellen@itf.fys.kuleuven.be}}
\affiliation{Institute for Theoretical Physics, University of Leuven, 3001 Leuven, Belgium}

\date{\today}

\begin{abstract}

One proposal for dS/CFT is that the Hartle-Hawking (HH) wave function in the large volume limit is equal to the partition function of a Euclidean CFT deformed by various operators. All saddle points defining the semiclassical HH wave function in cosmology have a representation in which their interior geometry is part of a Euclidean AdS domain wall with complex matter fields. We compute the wave functions of scalar and tensor perturbations around homogeneous isotropic complex saddle points, turning on single scalar field matter only. We compare their predictions for the spectra of CMB perturbations with those of a different dS/CFT proposal based on the analytic continuation of inflationary universes to real asymptotically AdS domain walls. We find the predictions of both bulk calculations agree to first order in the slow roll parameters, but there is a difference at higher order which, we argue, is a signature of the HH state of the fluctuations.

\end{abstract}

\maketitle

\section{Introduction}

One version of dS/CFT \cite{Strominger2001} conjectures that the wave function of the universe with asymptotic de Sitter (dS) boundary conditions is given in terms of the partition function of a Euclidean CFT deformed by certain operators. The wave function approach to dS/CFT was pioneered in \cite{Maldacena2002} and further explored in \cite{Garriga2008,Harlow2011,Maldacena2011,Castro2012,Banerjee2013,Anninos2011,Anninos2012,Anninos2014,Hertog2011,Hartle2012a,Hartle2012}. Other proposals for dS/CFT such as \cite{McFadden2009} rely on a procedure of analytic continuation to AdS. Both versions of dS/CFT are not the same. Here we study the differences between these two classes of proposals, at the semiclassical level, for the prediction of observables associated with primordial fluctuations in inflationary universes.

As a toy model for a wave function based proposal, we consider the semiclassical Hartle-Hawking (HH) or no-boundary wave function (NBWF) on $S^3$, in Einstein gravity coupled to a scalar with a positive potential and a positive cosmological constant with all other matter fields (or sources) set to zero. All saddle points defining the semiclassical NBWF have a representation in which their interior geometry is part of a Euclidean AdS domain wall \cite{Hertog2011}. In this `AdS representation' of the NBWF, its saddle points make a smooth transition from their approximate AdS regime to a Lorentzian (inflationary) universe that is asymptotically de Sitter \cite{Hertog2011}. The matter fields are in general complex along the AdS domain wall, because otherwise they wouldn't be both real at the final boundary and everywhere regular as the HH proposal prescribes. 

In the AdS representation, the tree level no-boundary measure on the classical ensemble of histories of the universe is specified by the regularised AdS domain wall actions. The transition region between AdS and dS merely compensates for the volume terms in the AdS action and accounts for the phases that explain the classical behaviour of the final configuration \cite{Hertog2011}. Hence by AdS/CFT the action of both the AdS regime and the transition region can be replaced by the logarithm of the partition function of a dual field theory. The asymptotic profile of the matter fields in the AdS regime enter as sources in the partition function. The dual no-boundary measure thus involves certain complex deformations of Euclidean CFTs familiar from AdS/CFT \cite{Hertog2011}.
The dependence of the partition function on the values of the sources, which are locally related to the argument of the asymptotic wave function, provides a holographic expression of the no-boundary measure.

In this paper we identify in detail the AdS representation for saddle points corresponding to homogeneous isotropic histories with an early period of scalar field driven inflation. Using this we compute the wave function of linear perturbations in the AdS representation, taking in account the effect of the imaginary component of the background saddle points.

We then contrast their predictions for the spectra of primordial perturbations with those of a different framework for dS/CFT, developed mostly by McFadden and Skenderis \cite{McFadden2009}, that is based on the analytic continuation of inflationary histories to real, Euclidean, AdS domain walls. The dual field theories in this model for dS/CFT are certain real deformations of the CFTs familiar from Euclidean AdS/CFT. This proposal does not directly specify the underlying quantum state in the cosmological domain, let alone the cosmological measure, but, assuming Bunch-Davies initial conditions, their framework can be used to predict the spectrum of primordial perturbations in inflationary universes from an AdS calculation or its dual \cite{McFadden2009,McFadden2010,McFadden2010b,McFadden2010c,McFadden2011,Bzowski2011,Bzowski}. 
We will find that the imaginary component of the scalar domain wall profile in the no-boundary saddle points only has a small effect on the spectrum of perturbations that leave the horizon during the slow roll phase of inflation. Specifically the difference is manifest at second order in the slow roll parameters. At the same time this suggests the differences between both approaches may well be more pronounced on the large, unobservable 
scales associated with eternal inflation. 

We argue that this difference can be traced to the underlying quantum state of the fluctuations. The complex structure of the NBWF saddle points specifies a state for the fluctuations that differs slightly from the exact Bunch-Davies initial conditions employed in \cite{McFadden2009}. This shows up in the detailed spectral properties of the primordial perturbation, which therefore provide an excellent probe of the quantum state of the fluctuations.

The paper is organized as follows: In Section \ref{rep} we review the holographic form of the semiclassical Hartle-Hawking wave function in Einstein gravity and the dS/CFT conjecture that is implied. In Section \ref{mini} we give an explicit realisation of the dS and the AdS representation of the wave function in the minisuperspace of homogeneous isotropic histories. In Section \ref{scalarpert} we compute the wave function of scalar perturbations using the AdS representation of the backgrounds and for a quadratic scalar potential. We perform a similar analysis for tensor fluctuations in Section \ref{tensorpert} and discuss in qualitative terms the generalisation of our conclusions to arbitrary single field models in Section \ref{genpot}. We conclude in Section \ref{conclusion} with a discussion of the difference between the NBWF predictions for perturbations and the predictions of dS/CFT proposals based on the analytic continuation to real AdS domain walls. Appendix A summarises useful formulae of cosmological perturbation theory.

\section{dS/CFT in the No-Boundary State}
\label{rep1}

We first review the connection between the NBWF with asymptotically de Sitter conditions and Euclidean AdS/CFT.

\subsection{The No-Boundary Wave Function}
\label{nbwf1}

We consider Einstein gravity coupled to a positive cosmological constant $\Lambda$ and a scalar field moving in a positive potential $V$ that has a regime where the conditions for slow roll inflation are satisfied. A quantum state of the universe in this model is specified by a wave function $\Psi[h_{ij},\chi]$ on the superspace of all three-metrics $h_{ij}(\vec x)$ and matter field configurations $\chi(\vec x)$ on a closed spacelike surface $\Sigma$. 

We assume the no-boundary wave function as a model of the state \cite{Hartle1983}. The NBWF is formally given by a sum over four-geometries $g$ and fields $\phi$ on a four-manifold $M$ with one boundary $\Sigma$. The contributing histories match the values $(h_{ij},\chi)$ on $\Sigma$ and are otherwise regular. They are weighted by $\exp(-I/\hbar)$ where $I[g,\phi]$ is the Euclidean action. Schematically, 
\begin{equation}  
\Psi[h_{ij}(\vec x),\chi (\vec x)] \equiv  \int_{\cal C} \delta g \delta \phi \exp(-I[g(x),\phi(x)]),
\label{nbwf2}
\end{equation}
where the Euclidean action $I[g(x),\phi(x)]$ is a sum of the Einstein-Hilbert action (in Planck units $\hbar=c=G=1$)
\begin{equation}
I_C[g] = -\frac{1}{16\pi}\int_M d^4 x (g)^{1/2}(R-2\Lambda) -\frac{1}{8\pi}\int_M d^4 x (h)^{1/2}K,
\label{curvact}
\end{equation}
and the matter action\footnote{We have chosen the normalization of the scalar field $\phi$ to simplify subsequent equations and maintain consistency with earlier papers, specifically \cite{Hartle2008}.} 
\begin{equation}
I_{\phi}[g,\Phi]=\frac{3}{4\pi} \int_M d^4x (g)^{1/2}\left[\frac{1}{2}(\nabla\phi)^2 +V(\phi)\right].
\label{mattact}
\end{equation}

In some regions of superspace the path integral \eqref{nbwf2} defining the NBWF can be approximated by the method of steepest descents. Then the NBWF will be approximately given by a sum of terms of the form 
\begin{equation}
\Psi[h_{ij},\chi] \approx  \exp(-I_R[h_{ij},\chi] +i S[h_{ij},\chi]) .
\label{semiclass}
\end{equation}
Here $I_R[h_{ij},\chi]$ and $-S[h_{ij},\chi]$ are the real and imaginary parts of the Euclidean action, evaluated at the saddle point. 
In regions of superspace where $S$ varies rapidly compared to $I_R$ (as measured by quantitative classicality conditions \cite{Hartle2008}) the NBWF predicts that the boundary configuration $(h_{ij},\chi)$ behaves classically. That is, with high probability the configuration will evolve according to classical deterministic laws in the no-boundary state. 

In the presence of a positive cosmological constant, the Wheeler-DeWitt equation implies that the classically conditions hold for general boundary configurations $(h_{ij},\chi)$ when the three-volume is sufficiently large \cite{Hartle2008,Hertog2011,Hartle2012}. In this regime the NBWF predicts an ensemble of spatially closed, Lorentzian asymptotically de Sitter histories.
Each individual classical history in this ensemble has a probability proportional to $\exp[-2 I_R(h_{ij},\chi)]$ to leading order in $\hbar$, which is conserved along the classical history as a consequence of the Wheeler-DeWitt equation (cf \cite{Hartle2008}). 

\subsection{Representations of complex saddle points}
\label{rep2}

We now discuss in more detail the saddle point geometries defining the semiclassical wave function. The line element of a closed three-geometry can be written as
\begin{equation}
d\Sigma^2 = b^2 \tilde h_{ij}(\vec{x})dx^idx^j,
\label{bdmetric}
\end{equation}
where $b$ is an overall scale factor and we take $\tilde{h}_{ij}(\vec{x})$ to have unit volume. Superspace is therefore spanned by $b$, $\tilde{h}_{ij}(\vec{x})$ and the boundary configuration $\chi (\vec{x})$ of the scalar field $\phi$. Thus $\Psi=\Psi(b,\tilde{h}_{ij},\chi)$. The compact saddle point geometries are of the form
\beq \label{sad}
ds^2=N^2(\lambda)d\lambda ^2+g_{ij}(\lambda,\vec{x})dx^idx^j,
\eeq
where $\{\lambda,x^i\}$ are real coordinates on the manifold $\mathcal{M}$. We take $\lambda =0$ to locate the South Pole (SP) of the compact saddle point, where the geometry smoothly caps off, and $\lambda=1$ to locate the boundary $\Sigma$ of $\mathcal{M}$. Regularity at the SP together with the boundary condition that geometry and field match $(b,\tilde{h}_{ij},\chi)$ on $\Sigma$ imply that the saddle points are generally complex solutions of the Einstein equation.

The Einstein equation can be solved for $\{g_{ij}(\lambda,\vec{x}), \phi(\lambda,\vec{x})\}$ for any complex $N(\lambda)$. Different choices of $N(\lambda)$ yield different geometric representations of the same saddle point. If we define the complex variable 
\begin{equation}
\tau(\lambda) \equiv \int_0^{\lambda} d \lambda' N(\lambda'),
\end{equation}
then different choices of $N(\lambda)$ correspond to different contours in the complex $\tau$-plane. Contours begin at the SP at $\lambda =0$ and end at the boundary at $\lambda=1$, where $\tau(1)\equiv \upsilon$. Each contour that connects the SP to $\upsilon$ yields a different representation of the same saddle point. This freedom in the choice of contour gives physical meaning to a process of analytic continuation --- not of the Lorentzian histories themselves as in the dS/CFT proposal of \cite{McFadden2009} --- but of the complex saddle points that define their probabilities in the no-boundary state. It was shown \cite{Hertog2011} that this freedom of contour can be used to identify two different useful representations of the saddle points corresponding to asymptotically de Sitter histories. In one representation (dS) the interior saddle point geometry behaves as if $\Lambda$ and $V$ were positive. In the other (AdS) the interior geometry behaves as if these quantities were negative, specifying a Euclidean approximate AdS regime. 

To make this more explicit consider the large volume expansion of the saddle points in terms of a variable $u$ defined by 
\be
u\equiv e^{iH\tau} \equiv e^{-Hy+iHx},
\label{defu}
\ee
with $H^2 \equiv \Lambda/3$. At large volume the general complex solution of the Einstein equation is 
\begin{align}
g_{ij}(u,\vec{x}) &= \frac{c^2}{u^2} \left[ \tilde{h}_{ij}^{(0)}(\vec{x})+ \tilde{h}_{ij}^{(2)}(\vec{x}) u^2  + \tilde{h}_{ij}^{(-)}(\vec{x}) u^{2\lambda_-}+ \cdots + \tilde{h}_{ij}^{(3)}(\vec{x}) u^{3} + \cdots \right], \label{a_series}\\
\phi(u,\vec{x}) &=  u^{\lambda _-}( \alpha(\vec{x}) + \alpha_1(\vec{x}) u + \cdots )   +  u^{\lambda _+}(\beta(\vec{x}) + \beta_1(\vec{x}) u + \cdots).
\label{phia}
\end{align}
where $\tilde{h}_{ij}^{(0)}(\vec{x})$ has unit volume, $\lpm \equiv (3/2)[1\pm q(m)]$ with $q(m) \equiv \sqrt{1-(2m/3H)^2}$ and we have assumed that $V$ is quadratic near its minimum.

In saddle points associated with asymptotically dS universes the phases of the fields at the SP are tuned so that $g_{ij}$ and $\phi$ become real for small $u$ along a vertical line $x= x_{dS}$ in the complex $\tau$-plane. Eqs. \eqref{a_series} - \eqref{phia} show that along this curve the complex saddle point tends to a real asymptotically dS history. The $x= x_{dS}$ line is part of a de Sitter contour from the SP to the boundary that we call $C_{dS}$. However, since the expansions are analytic functions of $u$, there is an alternative asymptotically vertical curve located at $x_{adS}=x_{dS}-\pi/(2H)$ along which the metric $g_{ij}$ is also real, but with the opposite signature. Along this curve the saddle point geometry \eqref{a_series} is asymptotically Euclidean AdS and the scalar moves in an effectively negative potential $-V$. One can thus envision a contour $C_{adS}$ which first approaches the $x=x_{adS}$ line and then cuts horizontally to the endpoint $\tau = \upsilon$. This contour has the same endpoint $\upsilon$, the same action, and makes the same predictions as $C_{dS}$. But the interior geometry is different; it consists of a regular Euclidean AdS domain wall that makes a smooth transition to an asymptotically dS universe. 

The action of a saddle point is an integral of its complex geometry and fields, which includes an integral over complex time $\tau$. Different contours for this time integral yield the same amplitude of the boundary configuration $(b,\tilde{h}_{ij},\chi)$. Thus we can use the AdS contour to compute the no-boundary measure. 
The real part of the Euclidean action along $x=x_{adS}$ has the usual AdS volume divergences for large $y$. By contrast the real part of the action is asymptotically constant along the $x=x_{dS}$ curve. Hence the contribution to the action from the horizontal branch of the AdS contour must regulate the divergences associated with the vertical part of $C_{adS}$. Indeed, adjusting for signature, it was shown in \cite{Hertog2011} that the divergent contributions to the action integral along the horizontal part of $C_{adS}$ are precisely the regulating counter terms $S_{ct}$ familiar from AdS/CFT, plus a universal phase factor $iS_{ct}$.  

By contrast, the action integral along the horizontal branch of $C_{adS}$ does not contribute to the asymptotically finite part of the action \cite{Hertog2011}. Hence the probabilities for Lorentzian asymptotically dS histories in the no-boundary state are fully specified by the regularised action of the interior AdS regime of the saddle points. In particular we have 
\be
I [\upsilon,\tilde h_{ij},\alpha(\vx)] = -I^{\rm reg}_{DW} [\tilde h_{ij},\tilde \alpha(\vx)] + iS_{ct}[\upsilon,\tilde h_{ij},\alpha(\vx)] +{\cal O}(e^{-Hy_\upsilon}) .
\label{scnbwf}
\ee
Here $I^{\rm reg}_{DW}$ is the $y \rightarrow \infty$ limit of the regulated AdS action of the domain wall and $\tilde \alpha \equiv \vert \alpha \vert e^{-i\lambda_{-}\pi/2}$ (cf eq.\eqref{phia}). The minus sign in front of $I^{\rm reg}_{DW}$ is connected to the fact that the NBWF behaves as a decaying wave function along the AdS branch of the contour \cite{Conti2015}. 

\subsection{Holographic No-Boundary Measure}
\label{rep}

The AdS representation of the saddle points provides a natural connection between the NBWF in the large volume limit and Euclidean AdS/CFT. In the supergravity approximation the Euclidean AdS/CFT dictionary states that
\be
\exp (-I^{reg}_{DW}[\tilde h_{ij},\tilde{\alpha]}) = Z_{QFT}[\tilde h_{ij},\tilde{\alpha}] \equiv \langle \exp \int d^3x \sqrt{\tilde h} \tilde{\alpha} {\cal O} \rangle_{QFT},
\label{operator}
\ee
where the dual QFT lives on the conformal boundary represented here by the three-metric $\th_{ij}$. For radial domain walls this is the round three-sphere, but in general $\tilde{\alpha}$ and $\tilde h_{ij}$ are arbitrary functions of all boundary coordinates $\vx$. 
Applying \eqref{operator} to \eqref{scnbwf} yields the following holographic form of the semiclassical NBWF at large volume \cite{Hertog2011},
\be
\label{dscft}
\Psi [b, \tilde h_{ij},\chi] = \frac{1}{Z_{QFT}[\tilde h_{ij}, \tilde \alpha] }\exp(iS_{ct}[b, \tilde h_{ij},\chi]).
\ee
 The probabilities of asymptotically dS histories in the no-boundary state are thus given by the inverse of the partition functions of certain complex deformations of AdS/CFT dual field theories defined on the boundary surface $\Sigma$. The arguments of the wave function enter as external sources $(\tilde h_{ij}, \tilde \alpha)$ in the partition function. The dependence of the field theory partition function on those sources then gives a holographic measure on asymptotically de Sitter configurations. 


A key difference between \eqref{dscft} and dS/CFT proposals based on the analytic continuation of Lorentzian histories is that in the latter case, the auxiliary Euclidean AdS space that enters is real. By contrast in \eqref{dscft}, as we have discussed, the scalar field profile along the AdS branch is complex. In the next sections we compute to what extent this difference affects the predictions for cosmological observables associated with primordial perturbations.

\section{Homogeneous Minisuperspace}
\label{mini}

In this section we develop in more detail the AdS representation of the no-boundary saddle points corresponding to homogeneous and isotropic boundary configurations, with $\tilde h_{ij}$ the metric of the round unit three sphere. We focus here on quadratic potentials $V(\phi) = \frac{1}{2} m^2 \phi^2$ and comment on the generalisation of our results to other potentials below in section \ref{genpot}. 

Homogeneous isotropic minisuperspace is spanned by the the scale factor $b$ and the homogeneous value $\chi$ of the scalar field on the boundary. The saddle points take the form 
\begin{equation}\label{homsad}
ds^2 = d\tau^2 + a^2(\tau) d \Omega^2_3,
\end{equation}
and the Euclidean action \eqref{curvact} and \eqref{mattact} reduces to
\begin{equation}
I[b, \chi] = \frac{3 \pi}{8} \int_0^v d \tau \left[-a \dot{a}^2 - a + H^2 a^3 + a^3 \dot{\phi}^2 + 2 a^3 V(\phi) \right],
\end{equation}
where the dot denotes a derivative with respect to $\tau$. The resulting equations of motion are
\begin{align}
\dot{a}^2 -1 + H^2 a^2 + a^2\left[-\dot{\phi}^2 + 2V \right] &= 0,  \label{bgeom1}\\
\ddot{\phi} + 3 \mathcal{H} \dot{\phi} - V_{,\phi} &= 0. \label{bgeom2}
\end{align}
Here $\mathcal{H} \equiv \dot{a}/a$ is the complex valued Hubble scale. In the large volume limit $|\mathcal{H}| = H$.

If $V$ has a regime where the slow roll conditions hold then there exists a one-parameter set of regular compact saddle points of the form \eqref{homsad}. The saddle points can be labeled by the absolute value $\phi_0$ of the scalar field at the SP. This is roughly equal to the value of $\phi$ at the start of inflation in the associated Lorentzian, inflationary history \cite{Hartle2008}.

Regularity implies that the behaviour of the saddle point solutions in the immediate neighbourhood of the SP is specified by
the potential term in \eqref{bgeom1}, yielding
\begin{equation}
\phi(\tau) \approx \phi(0), \qquad \qquad a(\tau) \approx \frac{\sin[m \phi(0) \tau]}{m \phi(0)}.
\label{norollsol}
\end{equation}
In the larger `inflationary' region around the SP the saddle points are approximately 
\begin{align}
\phi(\tau) &\approx \phi(0)+ i \frac{m \tau}{3}, &&a(\tau) \approx \frac{i}{2 m \phi(0)} \exp\left[-i m \phi(0) \tau + \frac{m^2 \tau^2}{6}\right], \notag \\
\dot{\phi}(\tau) &\approx i\frac{m}{3}, &&\mathcal{H}(\tau) \approx -i m \phi(0) + \frac{m^2 \tau}{3} = -3 \dot{\phi} \phi.
\label{LyonsdSsol}
\end{align}
Finally, at large scale factor when the field has rolled down and the solutions are dominated by the cosmological constant they are given by the asymptotic expansions \eqref{a_series} and \eqref{phia}.

\begin{figure}
\begin{center}
\includegraphics[scale=0.3]{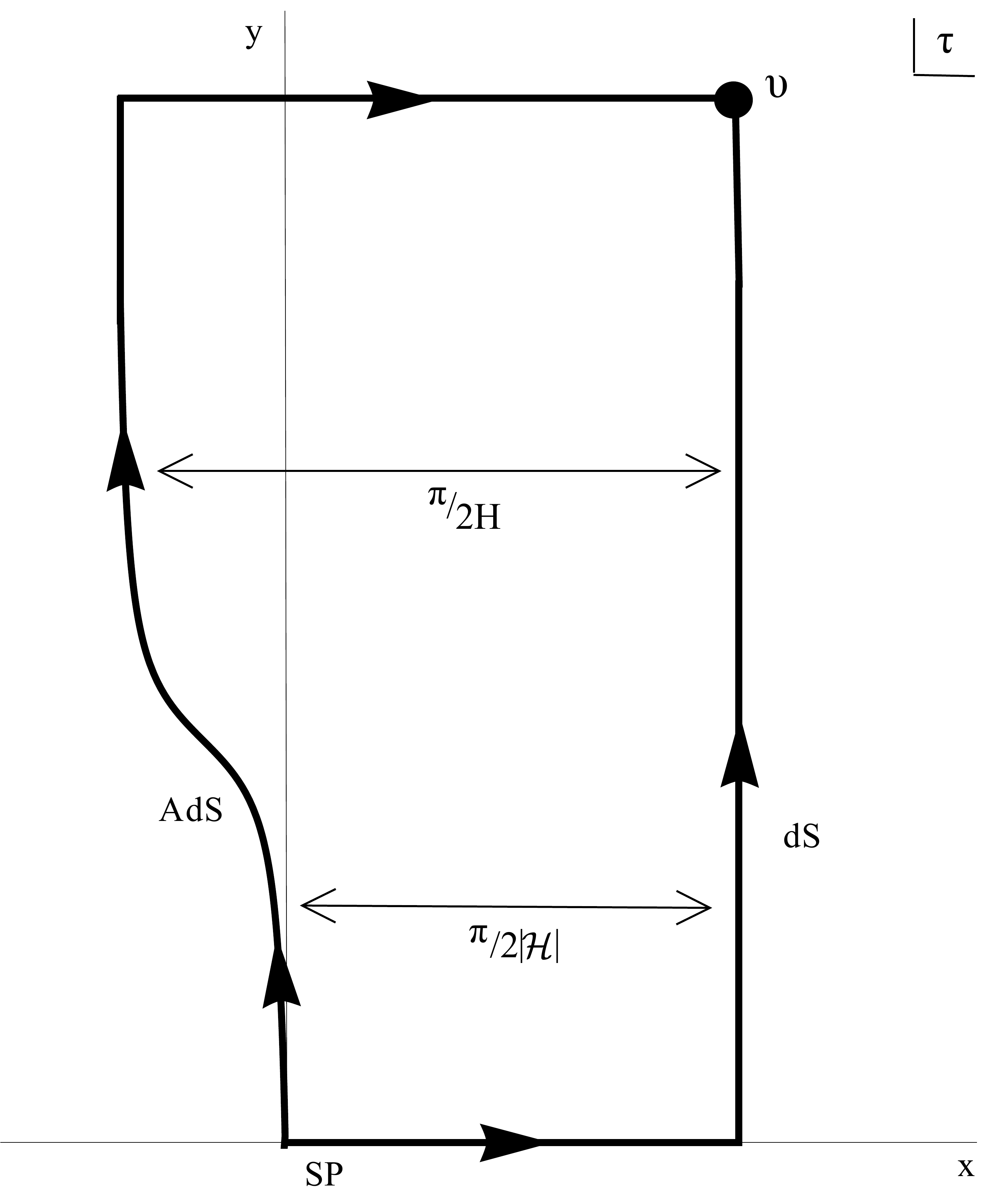}
\end{center}
\caption{The dS and AdS contours. {The dS contour runs from the SP along the x-axis to $x=x_{dS}$, at which point it runs vertically upward to the endpoint $\upsilon$. The AdS contour starts out vertically at the SP and gradually curves towards $\tau=x_{dS} - \pi/(2H) + i y_\upsilon$, from this point is runs horizontally to the endpoint $\upsilon$.}}
\label{contour}
\end{figure}

In saddle point solutions that correspond to asymptotic dS histories the phase of $\phi(0)$ at the SP is tuned so that both $a$ and $\phi$ become real along the vertical part of $C_{dS}$ at $x=x_{dS}$. This requires \cite{Lyons}
 \begin{equation}
x_{dS} = \frac{\pi}{2 m \phi_R(0)}, \qquad \qquad \phi_I(0) = -\frac{\pi}{6 \phi_R(0)},
\label{lyonsinitial}
\end{equation}
where $\phi_R (0)$ and $\phi_I(0)$ are the real and imaginary part of $\phi(0)$ respectively. Therefore the dS contour runs straight up from $x_{dS}$ in this regime, just as in the asymptotic regime discussed earlier. Numerical results \cite{Hartle2008} show that both regions actually smoothly join into a single vertical curve. Since $\partial_\tau = -i \partial_y$ along $x=x_{dS}$, it follows that $\dot{\phi}$ and $\mathcal{H}$ are pure imaginary. They can be related to the real Lorentzian quantities by,
\begin{equation}
\dot{\phi} = -i \phi'_L, \qquad \mathcal{H} = - i \mathcal{H}_L = 3i \phi \phi'_L, \label{L1}
\end{equation}
where prime denotes a derivative with respect to $y$. In the following all quantities will be Euclidean, unless indicated by the subscript $L$. To conclude, the dS representation of a saddle point is given by a contour in the complex $\tau$-plane that consists of two parts (cf. Fig \ref{contour}). The second part of the contour runs vertically upward from $(x_{dS},0)$ to the endpoint $(x_{dS},y_v)$. Along this part the geometry tends to that of a real, Lorentzian, asymptotic de Sitter space. The first part of the contour runs from the SP at $\tau=0$ along the x-axis to $x = x_{dS}$. Along this part of the contour the backgrounds are to a good approximation described by the no roll solutions \eqref{norollsol}. 

Along the AdS contour we have $a(\tau_{adS}) \simeq i a (\tau_{dS})$. From the solutions \eqref{LyonsdSsol} in the inflationary regime we find this holds for $x_{adS}(y) = x_{dS} - \pi/\left(2 |\mathcal{H}(y)|\right)$. In this regime $|\mathcal{H}(y)|= m \phi_{dS} (y)$, where $\phi_{dS}(y)$ is the value of the scalar field along the dS contour at the point $(x_{dS},y)$. For this value of $x_{adS}$ we have,
\begin{equation}
a_{adS}(y) = a_{dS}(y) \exp\left[\frac{i \pi}{2} + \frac{\pi^2}{24 \phi^2_{dS}(y)} \right],
\end{equation}
which for large $\phi$ indeed reduces to $a_{adS} \simeq i a_{dS}$. Along this contour the other quantities are related to their dS counterparts in the following way,
\begin{align}
\phi_{adS}(y) &= \phi_{dS}(y)\left[1 - \frac{i\pi}{6 \phi_{dS}^2(y)}\right], \label{phiAdS}\\
\dot{\phi}_{adS} &= \dot{\phi}_{dS},\\
\mathcal{H}_{adS}(y) &= \mathcal{H}_{dS}(y)\left[1 - \frac{i\pi}{6 \phi_{dS}^2(y)}\right] \label{HAdS}.
\end{align}
Note that in the limit of a small slow roll parameter $\epsilon \equiv |\dot{\phi}^2/\mathcal{H}^2|=1/(9\phi_{dS}^2)$ the scalar field is real and the Hubble scale imaginary along the AdS contour, just as along the dS contour. Thus in this limit the background approaches a real AdS domain wall. 

Since $x_{adS}$ depends on $y$ the contour will no longer go straight up. Instead it starts out vertically at the SP, in the no roll regime, where the scalar field is approximately real. But when the scalar field starts rolling down the potential the contour gradually moves away from the dS contour to the value $x_{dS}-\pi/(2H)$, which is reached when the cosmological constant starts to dominate.

We can thus deform the dS contour, while keeping the endpoints fixed, into a contour with an intermediate regime where the geometry is approximately Euclidean AdS. This is illustrated in Fig \ref{contour}, where the AdS regime of the solutions corresponds to the vertical part of the AdS contour to the turning point $\tau = x_{adS}(y_h)+ i y_h$. As discussed above the contribution to the action from the horizontal part of the contour regulates the AdS volume divergences and provides the universal phase factor of the wave function.

\section{Wave Function of Perturbations}
\label{pert}

We now turn to the wave function of perturbations. We consider linearized perturbations away from closed, homogeneous and isotropic three-geometries and field configurations. The extended minisuperspace is thus spanned by the scale factor $b$ of the homogeneous three-geometries, the homogeneous value of the scalar field $\chi$ and the parameters defining the perturbation modes. We denote the latter collectively by $z=(z_1,z_2,...)$ and $t_{ij}=(t_{ij}^1,t_{ij}^2,...)$ and define these precisely below.
Thus, $\Psi=\Psi(b,\chi, z,t_{ij})$.  

The NBWF is an integral \eqref{nbwf2} of the exponential of minus the Euclidean action $I$ over complex four-geometries and field configurations$(a(\tau),\phi(\tau),\zeta(\tau),\gamma_{ij}(\tau))$ that are regular on a four-disk with a three-sphere boundary on which the four-dimensional histories take the real values $(b,\chi,z,t_{ij})$ \cite{Hartle1983,Halliwell}. 

We restrict to linear fluctuations by only retaining up to quadratic terms in the perturbations in the action in \eqref{nbwf2}:
\begin{equation}
I = I^{(0)}[a(\tau),\phi(\tau)] + I^\epb [a(\tau),\phi(\tau),\zeta(\tau),\gamma_{ij}(\tau)]  .
\label{pertaction}
\end{equation}
Then $I^{(0)}$ describes the homogeneous isotropic background and $I^\epb$ describes the quadratic perturbations away from that background. The linear term vanishes due to the field equations.

In regions of superspace where the integral in \eqref{nbwf2} over $a(\tau)$ and $\phi(\tau)$ defining the homogeneous background can be approximated by the method of steepest descents the wave function $\Psi$ will be of the form
\begin{equation}
\Psi(b,\chi,z,t_{ij}) \approx \exp\left[-I^{\epa}_R(b,\chi) + i S^{\epa}(b,\chi)\right] \psi(b,\chi,z,t_{ij}),
\label{semiclassback}
\end{equation}
where the perturbation wave function $\psi$ is defined by the remaining integral over $\zeta$ and $\gamma_{ij}$,
\begin{equation}  
\psi(b,\chi,z,t_{ij}) \equiv  \int_{\cal C}\delta\zeta \delta \gamma  \exp\left(-I^\epb[a(\tau),\phi(\tau),\zeta(\tau),\gamma_{ij}(\tau)]\right) .
\label{qftwf}
\end{equation}
When evaluated on one of the classical background histories the wave function \eqref{qftwf} becomes a function of time. The Wheeler-DeWitt equation then implies a Schr\"odinger equation for $\psi(z,t_{ij})$ that describes the evolution of the state of the fluctuations in the background $(b(t),\chi(t))$. In this way, the fluctuation fields can be thought of as quantum fields on the possible background classical spacetimes, with the state of the fields determined by the NBWF through \eqref{qftwf}. 

The integral defining the wave function in \eqref{qftwf} may itself be approximated by the method of steepest descents. The result for a particular extremum $a(\tau), \phi(\tau)$ of $I^\epa$ is
\begin{equation}
\psi(b,\chi,z,t_{ij}) \propto  \exp\left[-I^{\epb}_R(b,\chi,z,t_{ij}) + i S^{\epb}(b,\chi,z,t_{ij})\right] .
\label{semiclassfluct}
\end{equation}
The extremizing histories $\zeta(\tau)$ and $\gamma_{ij}(\tau)$ are regular on the manifold of integration and match $z$ and $t_{ij}$ at its one boundary. $I^{\epb}_R(b,\chi,z,t_{ij})$ and $-S^{\epb}(b,\chi,z,t_{ij})$ are the real and imaginary parts of the action $I^\epb$ evaluated on this history.  

As for the backgrounds, this quantum mechanical theory of fluctuations around a classical background universe will predict they behave classically in regions of superspace where $S^\epb(b,\chi,z,t_{ij})$ varies rapidly in $z$ and $t_{ij}$ compared to $I^\epb(b,\chi,z,t_{ij})$ \cite{Hartle2008,Hartle2010}. Specifically when this is the case the wave function \eqref{semiclassfluct} predicts an ensemble of suitably coarse grained, classical, perturbed histories $z(t)$ and $t_{ij}(t)$ that with high probability lie along the integral curves of $S^\epb(b,\chi,z,t_{ij})$. The probabilities of the classical fluctuations in a given homogeneous isotropic background are then proportional to  $\exp[-2 I^{(2)}(b,\chi,z,t_{ij})]$.

\subsection{Scalar perturbations}
\label{scalarpert}

We first consider scalar perturbations. The wave function of linear scalar fluctuations around the homogeneous isotropic histories predicted by the NBWF was computed in \cite{Halliwell,Hawking1993,Hartle2010} using the dS representation of the background saddle points. Here we briefly review this result. Then we perform the analogue calculation in the AdS representation of the saddle points and compare this with the result in other approaches to dS/CFT.

The general perturbed metric can be written as
\begin{align}
ds^2 &=(1+2\varphi)d\tau^2 + 2a(\tau) B_{|i}dx^i d\tau + a^2(\tau)[(1-2\psi)\bar \gamma_{ij} + 2 E_{|ij}]dx^i dx^j,
\end{align}
where $\bar \gamma_{ij}$ is the metric of the unit radius three-sphere, $x^i$ the coordinates on the three sphere and the vertical bar denotes covariant differentiation with respect to $\bar \gamma_{ij}$. Expanding the perturbations in the standard scalar harmonics $Q^{n}_{lm} (x^i)$ on $S^3$ gives the definitions
\begin{align}
\varphi &= \frac{1}{\sqrt{6}} \sum_{nlm} g_{nlm} Q^n_{lm},  &&B = \frac{1}{\sqrt{6}} \sum_{nlm} \frac{k_{nlm} Q^n_{lm}}{(n^2-1)}, \notag \\
\psi &= \frac{-1}{\sqrt{6}} \sum_{nlm} (a_{nlm} + b_{nlm})Q^n_{lm},  &&E = \frac{1}{\sqrt{6}} \sum_{nlm} \frac{3 b_{nlm} Q^n_{lm}}{(n^2-1)}.
\label{metricexp}
\end{align}
and 
\be
\delta \phi (\tau, x) = \frac{1}{\sqrt{6}}  \sum_{nlm} f_{nlm} Q^{n}_{lm}.
\label{scalarexp}
\ee
Denoting the labels $n,\ l,\ m$ collectively by $(n)$ we write the expansion coefficients as $a_{(n)}, b_{(n)}, f_{(n)}, g_{(n)},k_{(n)}$.

There are five scalar degrees of freedom. However $g_{(n)}$ and $k_{(n)}$ appear as Lagrange multipliers in the action. Variations of the action with respect to those variables give the linear Hamiltonian and momentum constraints. The NBWF satisfies the operator form of these constraints so it depends only on the background variables $b$ and $\chi$, and on a single linear combination of the boundary values of the perturbation variables $a_{(n)}, b_{(n)}, f_{(n)}$ -- the three functions that describe the perturbed three geometry. One can take this combination to be the following,
\be
\zeta_{(n)} = a_{(n)}+b_{(n)} - \frac{\mathcal{H}}{\dot \phi} f_{(n)},
\label{zet}
\ee
Hence one has $\psi (b,\chi,z)$, where $z \equiv (z_{(1)},z_{(2)},...)$ are the real values of $\zeta=(\zeta_{(1)},\zeta_{(2)},...)$ at the boundary. 

To first order in perturbation theory the semiclassical wave function \eqref{semiclassfluct} takes the form
\be
\psi (b,\chi,z) = \prod_{(n)} \psi_{(n)} (b,\chi,z_{(n)}), 
\label{swave}
\ee
with $I^{(2)}_{(n)}[b,\chi,z_{(n)}]$ the action of each mode, whose explicit form can be found in appendix \ref{appendix}. For perturbation modes that leave the Hubble radius during inflation it can be found by solving the complex perturbation equations in the slow roll backgrounds \eqref{LyonsdSsol}. Regularity at the SP implies that to leading order in $\tau$ one has $\zeta_{(n)} = \zeta_{(n)} (0) \tau^{n}$, where $\zeta_{(n)}(0)\equiv \vert \zeta_{(n)}(0) \vert  e^{i\theta} \equiv \zeta_{(n)0} e^{i\theta}$ is a complex constant. Its phase $\theta$ should be fine-tuned such that $\zeta_{(n)}$ is real at the boundary, and its amplitude $\zeta_{(n)0}$ is determined by the value of the boundary perturbation $z_{(n)}$. 

At small $\tau$ the wavelength $|a/n|$ of the perturbation modes is shorter than the horizon size since $|a\mathcal{H}| \rightarrow 1$ when $\tau \rightarrow 0$. In this regime the complex solution for $\zeta_{(n)}$ oscillates and is independent of the nature of the potential. On the other hand at larger $\tau$, when $n\ll |a\mathcal{H}|$, the general perturbation is a combination of a constant and a decaying mode. Thus one expects the wave function $\psi_{(n)} (b,\chi,z_{(n)})$ will depend only on the behavior of the potential for values of $\phi$ near the value taken by $\phi (\tau)$ at the time the perturbation leaves the horizon. The requirement that $\zeta_{(n)}$ is real at the boundary means that the phase $\theta$ of $\zeta_{(n)} (0)$ at the SP should be tuned such that the imaginary component of the subhorizon mode function matches onto the decaying mode at horizon crossing. It turns out this implies that a perturbation mode will become classical when its physical wavelength becomes much larger than the Hubble radius. The real part of the Euclidean action $I^{(2)}_{(n)}[b,\chi,z_{(n)}]$ tends to the following constant when the mode leaves the horizon, 
\be
\label{conserved}
I^{(2)}_{(n)}[b,\chi,z_{(n)}] =   \frac{\epsilon_{*}}{2\mathcal{H}^2_{L*}} n^3 z_{(n)}^2,
\ee
where $\epsilon \equiv \dot \chi^2/\mathcal{H}^2_L$ is the slow-roll parameter. The subscript $*$ on a quantity in \eqref{conserved} means it is evaluated at horizon crossing during inflation. Equation \eqref{conserved} specifies the probabilities of linear, classical perturbations around the homogeneous isotropic, inflating histories predicted by the NBWF. One sees that the probabilities of $z_{(n)}n^3$ are Gaussian, with variance $\mathcal{H}^2_{L*}/\epsilon_{*}$ characteristic of inflationary perturbations. 

We now evaluate the wave function of perturbations using the AdS representation of the saddle points in which the backgrounds are Euclidean AdS domain walls. The complex perturbation solutions that enter in the calculation are the same as before, because they are fully determined by regularity at the SP and the requirement that the perturbation be real at the boundary. However we are now interested in the solutions along the AdS contour. We start at small $\tau$, when the wavelength $|a/n|$ of the perturbation modes is shorter than the horizon size. To leading order in both $n/a$ and $\phi$ the regular solution is given by,
\begin{equation}
\zeta_{(n),adS}^{\text{in}} = -\zeta_{(n)} (0) \frac{\mathcal{H}_{adS}}{a_{adS} \dot{\phi}}  e^{n \eta_{adS}},
\end{equation}
where $\eta$ is the conformal time defined by $d\tau = a d\eta$ and $\zeta_n(0)$  is the same complex constant as above. Along the AdS contour $\eta_{adS}$ is approximately real, leading to a growing behaviour of the solution.
The regularised AdS action for the modes does not stabilise as long as the wavelength is shorter than the horizon scale.
However, the AdS scale factor increases along the vertical part of the AdS contour, and at some point the wavelength will become larger than the local Hubble radius. In this regime the perturbation solutions along the AdS contour consist of a constant and a decaying term, 
\begin{equation}
\zeta_{(n)}^g = c_n^g, \qquad \zeta_{(n)}^d = \frac{c^d_n \phi}{a^3},
\label{outhorexp}
\end{equation} 
where $c_n^g$ and $c_n^d$ are constants\footnote{The solutions are to leading order in $\phi$, the next order contains a term $ c_n/(a \phi \dot{\phi})^{2}$. Here $c_n$ is a constant that solely depends on $c_n^g$ and $n$, and is therefore also real. Since this term is lower order in $\phi$ it does not contribute to the constant part of the action. We can therefore ignore it in the analytic analysis of the action, but it will be relevant in our numerical analysis of the AdS contour below.}. The boundary conditions imply that $c^g_n$ is real and equal to $z_{(n)}$. The value of $c^d_n$ is not determined by the boundary conditions, but depends on the full evolution from the SP. We use numerical simulations to find its value. The AdS contour is not a straight line, but we simulate the solution first along the vertical curve $\tau = i y$ and then we discuss and quantify the corrections due to the deviation of the exact AdS contour from this.

\begin{figure}
\begin{tabular}{cc}
\includegraphics[scale=0.59]{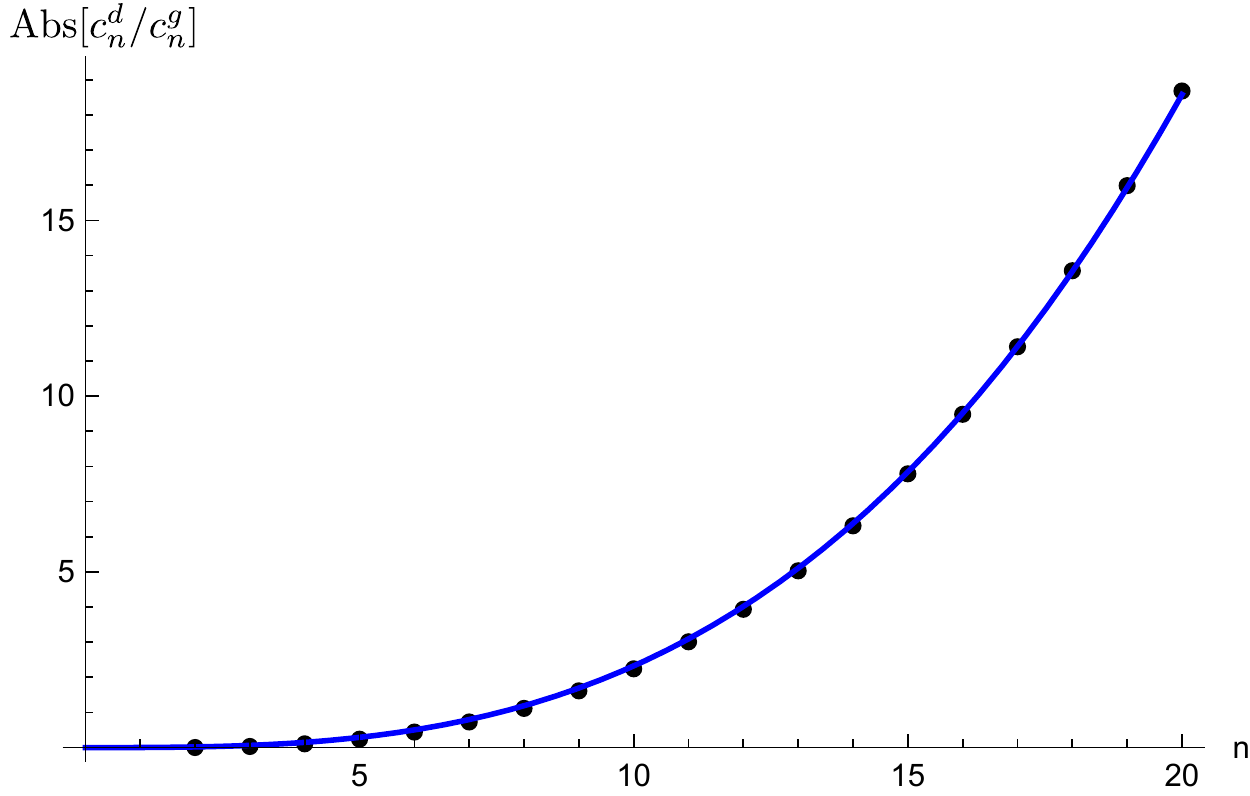}&
\includegraphics[scale=0.59]{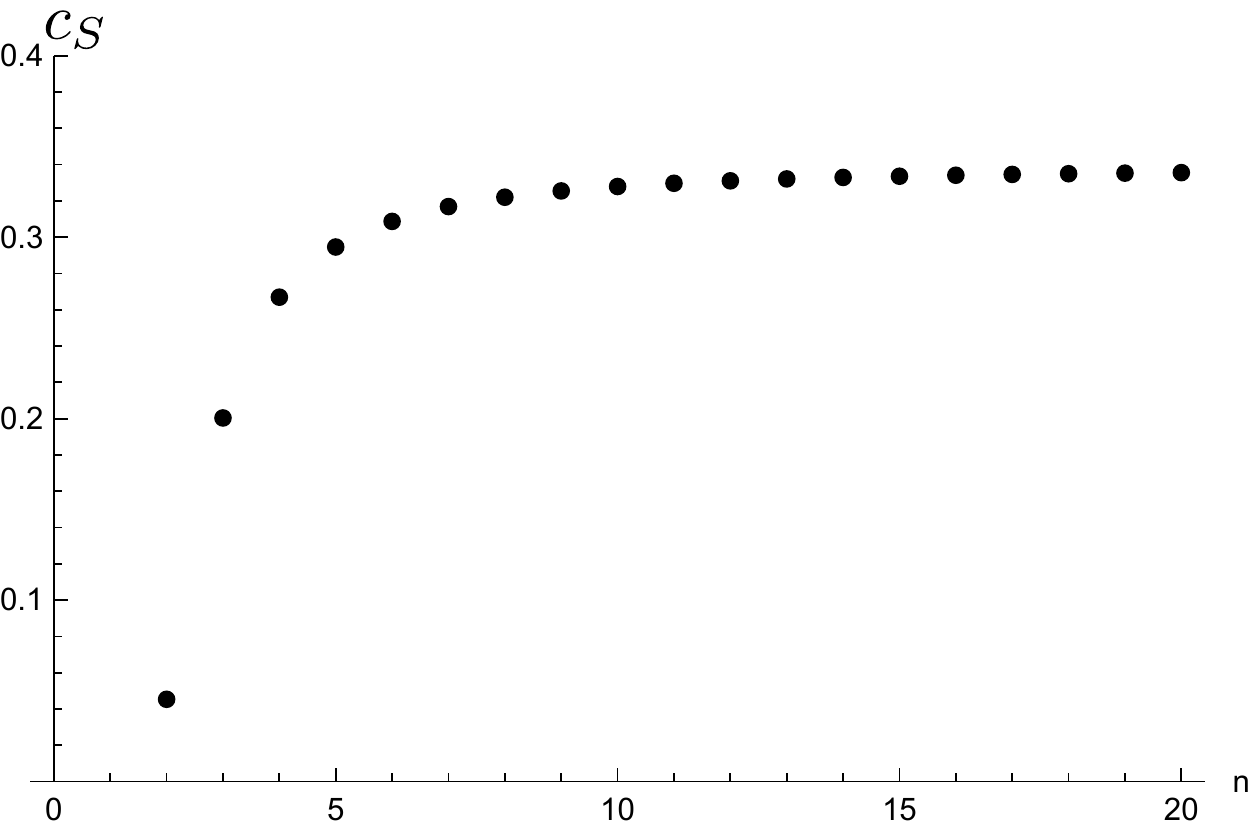}\\
\includegraphics[scale=0.59]{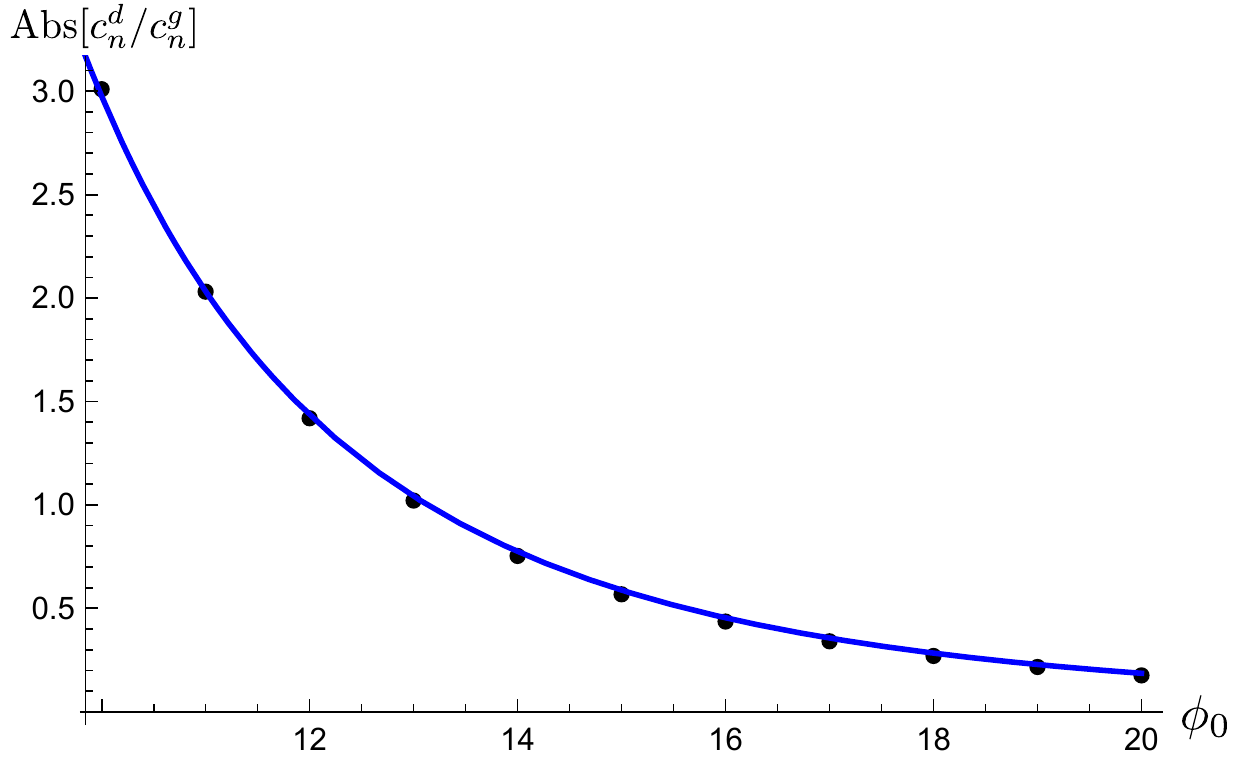}&
\includegraphics[scale=0.59]{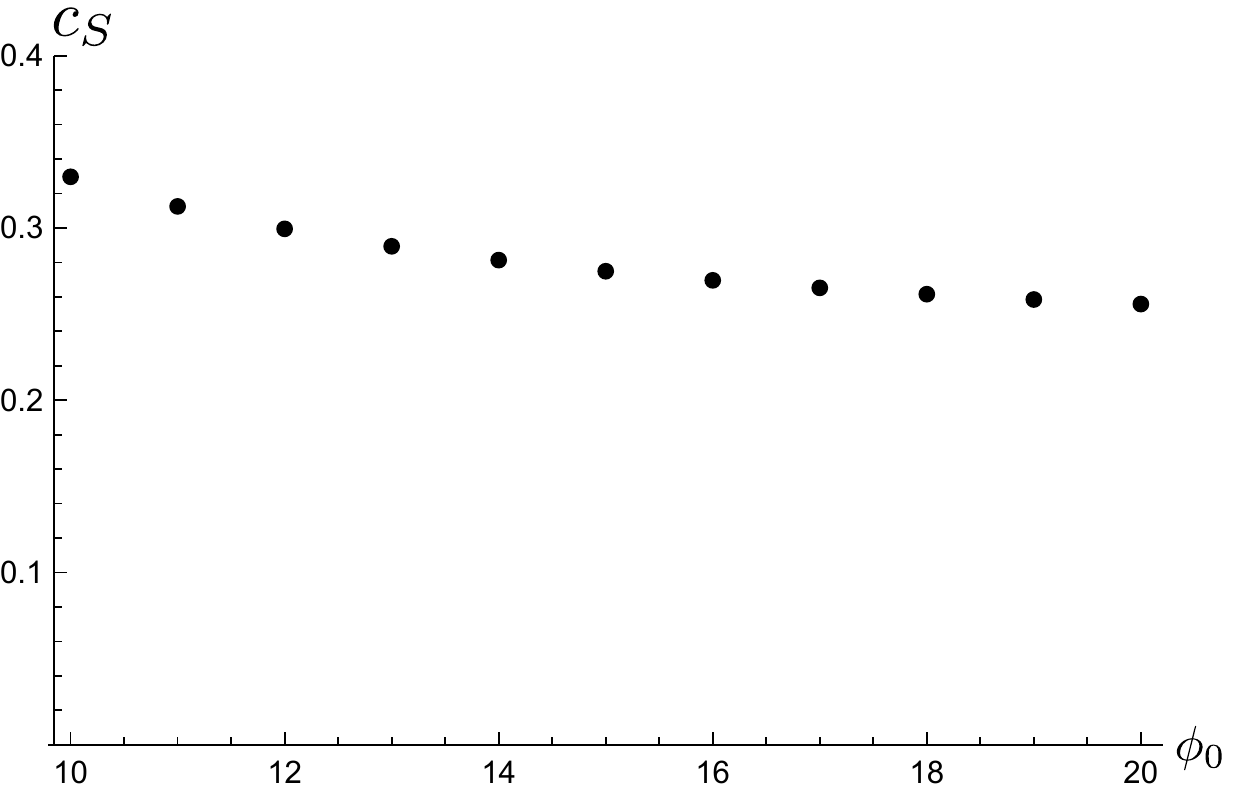}
\end{tabular}
\caption{The upper left plot shows the numerical results for Abs$[c^d_n/c^g_n]$ as a function of $n$, with $m^2 = 0.05$, $\phi_0=10$, $\theta_{\phi_0}= - \pi/(6 \phi_0^2)$ and $|\zeta(0)|=0.01$.  The solid line shows $\text{Abs}[c^d_n/c^g_n] \propto n^3$. The lower left plot shows Abs$[c^d_n/c^g_n]$ as a function of $\phi_0$, with $m^2 = 0.05$, $n=11$, $\theta_{\phi_0}= -\pi/(6 \phi_0^2)$ and $|\zeta(0)|=0.01$.  The solid line shows Abs$[c^d_n/c^g_n] \propto \phi(0)^{-4}$. The combined result gives Abs$[c^d_n/c^g_n] = c_S\, b^3_*/\chi_*$. The numerical values for the proportionality constant $c_S$ are shown in the right plots for different $n$ and $\phi_0$.}
\label{fign}
\end{figure}

\begin{figure}
\begin{center}
\includegraphics[scale=0.7]{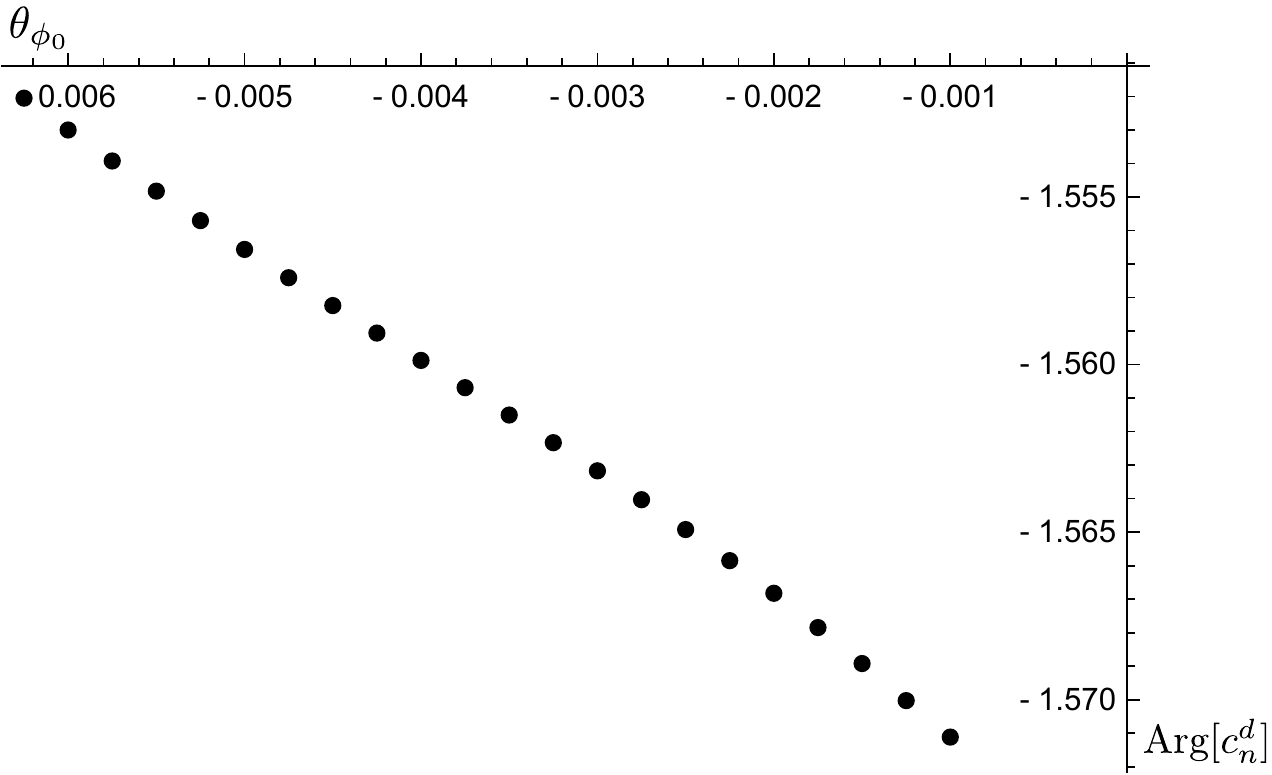}
\end{center}
\caption{Numerical results for the phase of the constant $c^d_n$ as a function of the initial phase of $\phi(0)$, for $m^2 = 0.05$, $n=15$ and $|\zeta(0)|=1$. The absolute value of $\phi(0)$ is determined by \eqref{lyonsinitial}.  The plot shows that $\text{Arg}[c^d_n] \simeq -\pi/2 -3.5\, \theta_{\phi_0}$.}
\label{figcangle}
\end{figure}

We simulated the evolution of $\zeta_{(n)}$ for different values of $n$, $\phi_0$ and $\theta_{\phi_0} \equiv \text{Arg}[\phi(0)]$. The numerical results are then matched onto the asymptotic analytic result \eqref{outhorexp} to find the dependence of $c_n^d$ on these variables. The results can be found in figures \ref{fign} and \ref{figcangle}. The left plots in figure \ref{fign} show that the ratio $c^d_n/c^g_n \propto n^3/\phi_0^4 \propto b^3_*/\chi_*$. Here we used the fact that $n$ and $\phi_0$ are related to the superspace coordinates by $n = b_* \mathcal{H}_{L*}$ and $\phi_0\propto \chi_*$. The absolute value of the proportionality constant is shown in the right plots of figure \ref{fign}\footnote{The plots show a  deviation from the constant value $1/3$ for low $n$ and large $\phi_0$. This is mainly due to ignoring higher order corrections in $\phi$ in the expansion \eqref{outhorexp} and the fact that the background is not yet fully classical for these values.}, while its phase can be seen in figure \ref{figcangle}. The final result is,
\begin{equation}
\frac{c^d_n}{c^g_n} = - \frac{i e^{-3.5i \theta_{\phi_0}} b^3_* }{3 \chi_*} .
\label{cdcgfrac}
\end{equation}
The phase $\theta_{\phi_0}$ in \eqref{cdcgfrac} is related to the superspace coordinates by
\begin{equation}
\theta_{\phi_0} \simeq - \frac{ \pi}{6 \chi^2_*} = - \frac{3 \pi}{2} \epsilon_*,
\end{equation}
which can be derived from equation \eqref{lyonsinitial}. Note that the stronger the slow roll conditions are satisfied, the smaller the phase.

In the above we made the approximation of a straight AdS contour. We now discuss the corrections due to the bending of the contour. 
{In the regime where the cosmological constant is sub-dominant it follows from \eqref{LyonsdSsol} that along the AdS contour}
\begin{equation}
\frac{d}{d\tau} =  -i e^{- \frac{3\pi i}{2}  \epsilon_*}  \frac{d}{dy}.
\label{dtau}
\end{equation}
This correction is of the exact same order as what we found for the complex scalar field above. It affects the action $I^{(2)}_{(n)}$ in two different ways. First of all it gives an extra phase to $c^d_n$. Unfortunately it is not possible to find the exact magnitude of this correction without doing the explicit simulation along the bent contour, which is beyond the scope of this work. Secondly the bending of the contour leads to an additional phase in each explicit derivative term in the action of the form $\exp[-3i \pi \epsilon_*/2]$.

Combining all this, therefore, means the constant part is of the form 
\begin{equation}
I_{(n),AdS}^{const} = \frac{a^3 \dot{\phi}^2}{2 \mathcal{H}^2} \dot{z}_{(n)} z_{(n)} = e^{c i  \pi \epsilon_* } \frac{ \epsilon_*}{2 \mathcal{H}^2_{L*}} n^3 z_{(n)}^2,
\label{AdSaction}
\end{equation}
where $c$ is a number of order 1. The overall phase gives a contribution to the real constant part of the action of order $ \mathcal{O}(\epsilon_*^{-2})$. We therefore conclude that
\begin{equation}
 I_{(n)}^{(2)} = -I_{(n),DW}^{(2),reg}
\end{equation}
up to slow roll corrections, in accordance with the general result \eqref{scnbwf}.

\subsection{Tensor perturbations}
\label{tensorpert}

Finally we evaluate the wave function of the tensor perturbations $t_{ij}$ in the AdS representation. To quadratic order in the action these decouple from the scalar modes. To solve for the wave function we perform a steepest descents approximation to the integral \eqref{qftwf} over $\gamma_{ij}$. Expanding the perturbations in the  transverse traceless tensorial harmonics $G^n_{ij}(x)$ on $S^3$ yields the following equation for the evolution of the complex expansion coefficients $d_{(n)} (\tau)$ 
\begin{equation}
\ddot{d}_{(n)} + 3 \mathcal{H} \dot{d}_{(n)} - (n^2-1)a^{-2}d_{(n)} = 0.
\label{tensoreq}
\end{equation}
As for the scalar perturbations the contribution to the wave function of each mode decouples and is just a boundary term \cite{Halliwell}, 
\begin{equation}
I_{(n)}^{(2)} = \frac{1}{2}a^3 t_{(n)} \dot{t}_{(n)} + 2 a^3 \mathcal{H} t_{(n)}^2,
\label{tensoraction}
\end{equation}
Here the dot denotes a derivative with respect to $\tau$, perpendicular to the boundary. 
We will evaluate this action in the AdS representation. 

The general asymptotic solution of \eqref{tensoreq} - when the modes are well outside the horizon - is given by
\begin{equation}
d_{(n)} = d^g_n - \frac{d^g_n(n^2-1)}{2 \mathcal{H}^2 a^2} + \frac{d^d_n}{a^3} + ...,
\label{tensorsol}
\end{equation}
were $d^g_n$ and $d^d_n$ are constants. The boundary value $d^g_n$ must be real and equal to $t_{(n)}$. To determine the value of the constant $d^d_n$ along the AdS contour we perform a numerical simulation.

\begin{figure}
\begin{center}
\begin{tabular}{cc}
\includegraphics[scale=0.59]{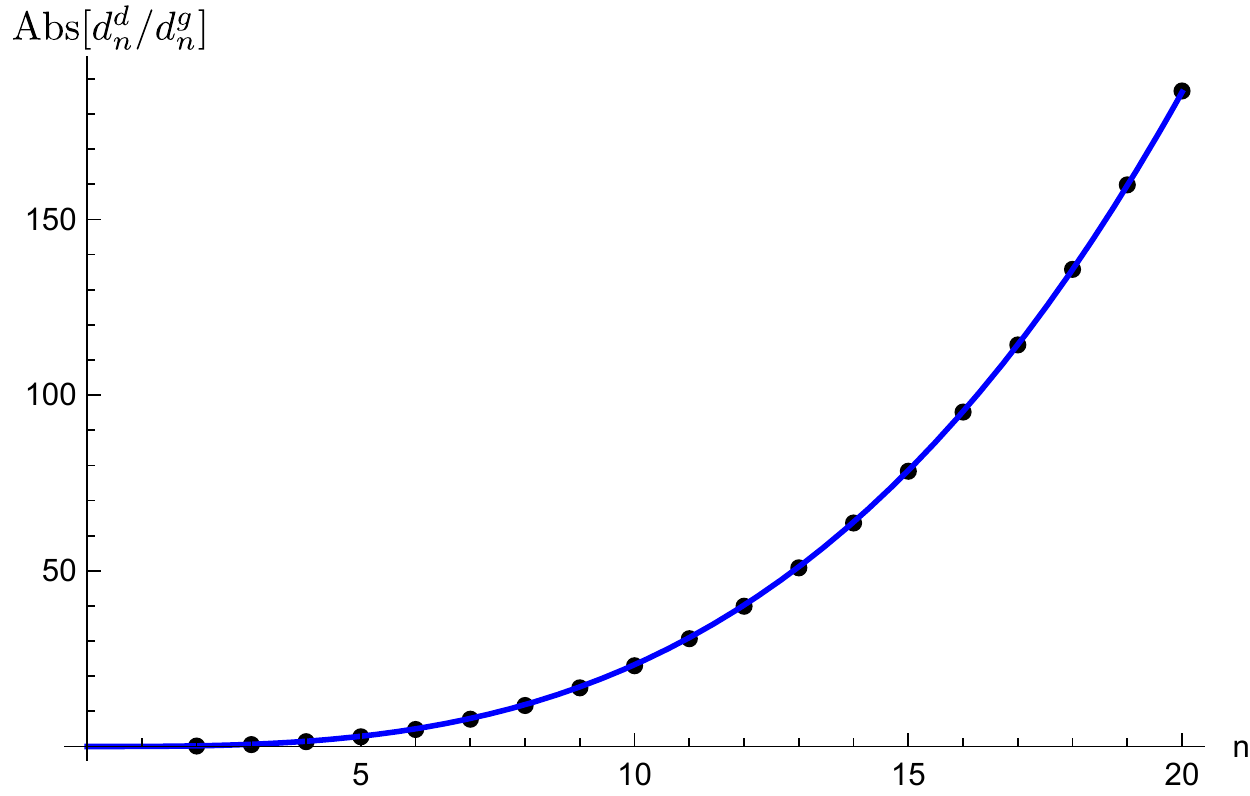}&
\includegraphics[scale=0.59]{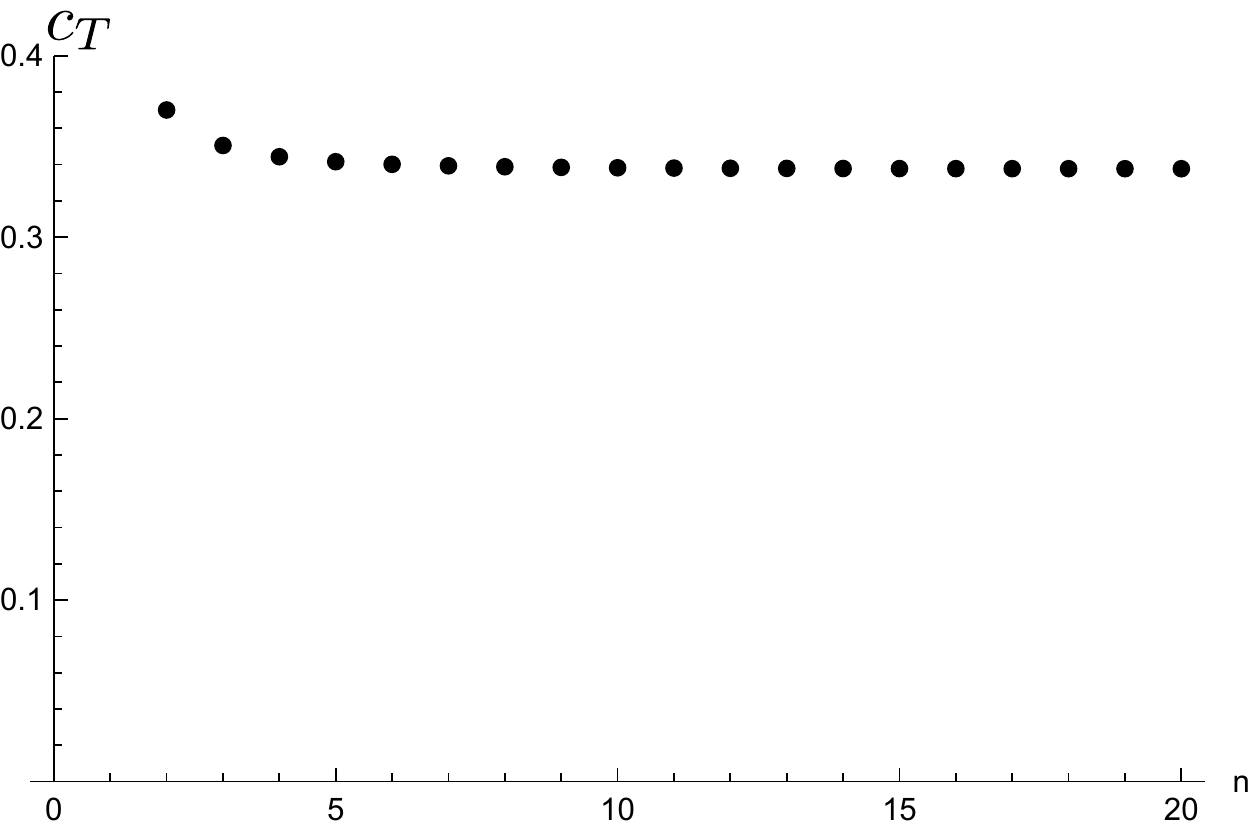} \\
\includegraphics[scale=0.59]{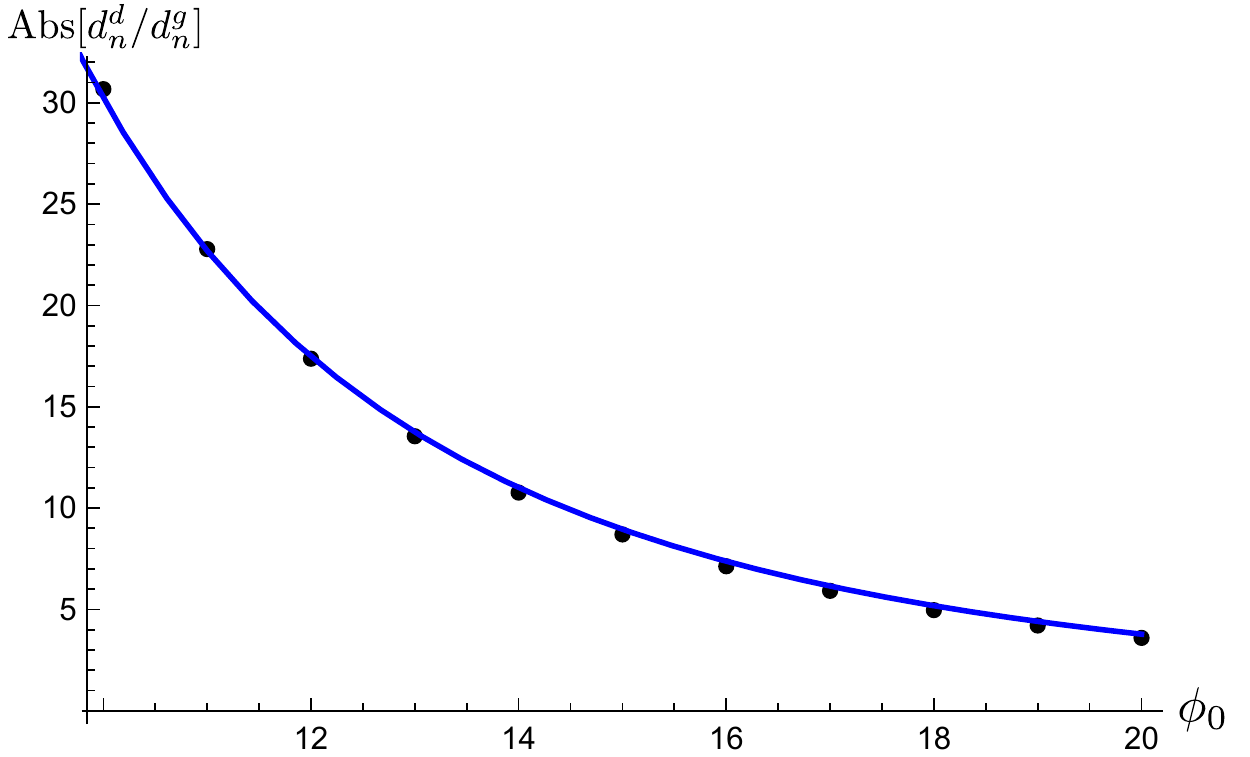}&
\includegraphics[scale=0.59]{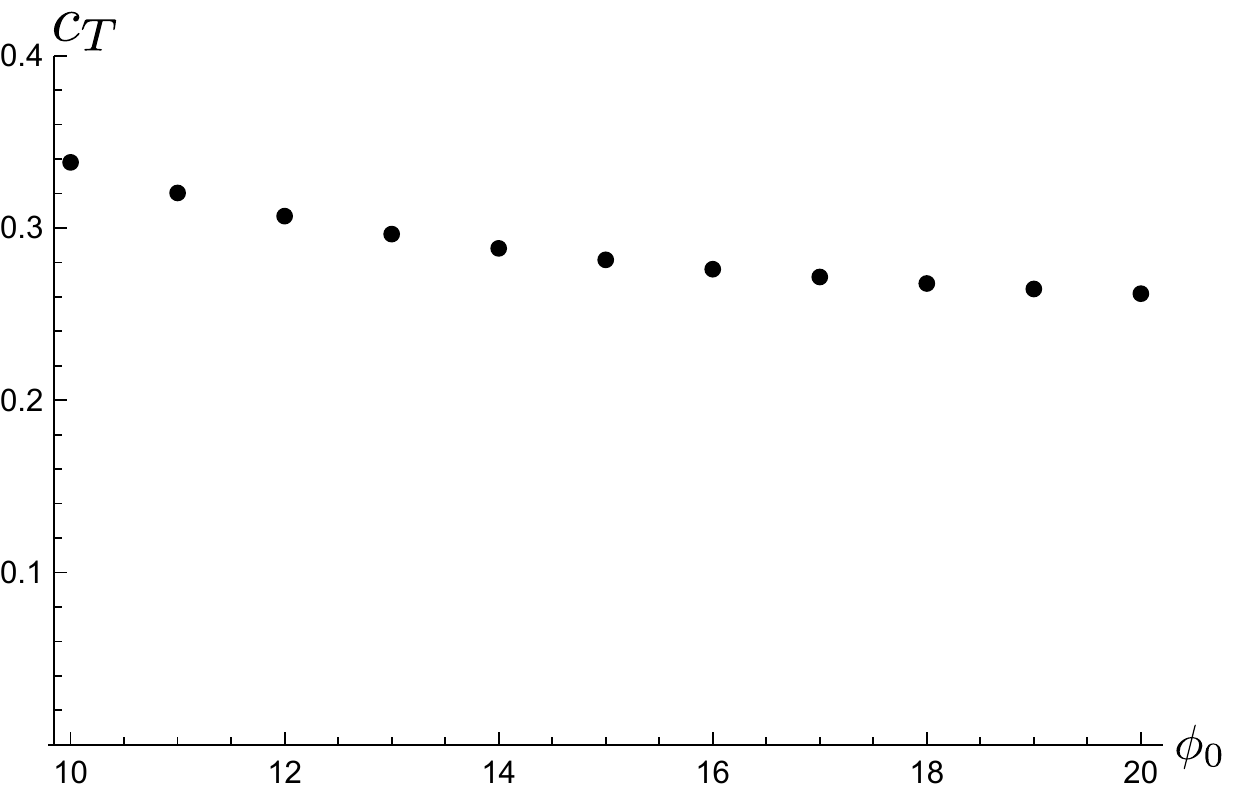} 
\end{tabular}
\end{center}
\caption{The upper left plot shows Abs$[d^d_n/d^g_n]$ as a function of $n$, with $m^2 = 0.05$, $\phi_0=10$, $\theta_{\phi_0} = - \pi /(6 \phi_0^2)$ and $|d_n(0)|=1$, where $|d_n(0)|$ is the initial amplitude of the tensor mode.  The solid line shows Abs$[d^d_n/d^g_n] \propto n^3$. The lower left plot shows Abs$[d^d_n/d^g_n]$ as a function of $\phi_0$, for the same parameters and $n=1$.  The solid line shows Abs$[d^d_n/d^g_n] \propto \phi_0^{-3}$. The numerical simulations indicate that Abs$[d^d_n/d^g_n] =  c_T\;b^3_*$. The numerical values for the constant $c_T$ are shown in the right upper and lower plot for different $n$ and $\phi_0$ respectively.}
\label{tensorsim}
\end{figure}

\begin{figure}
\begin{center}
\includegraphics[scale=0.7]{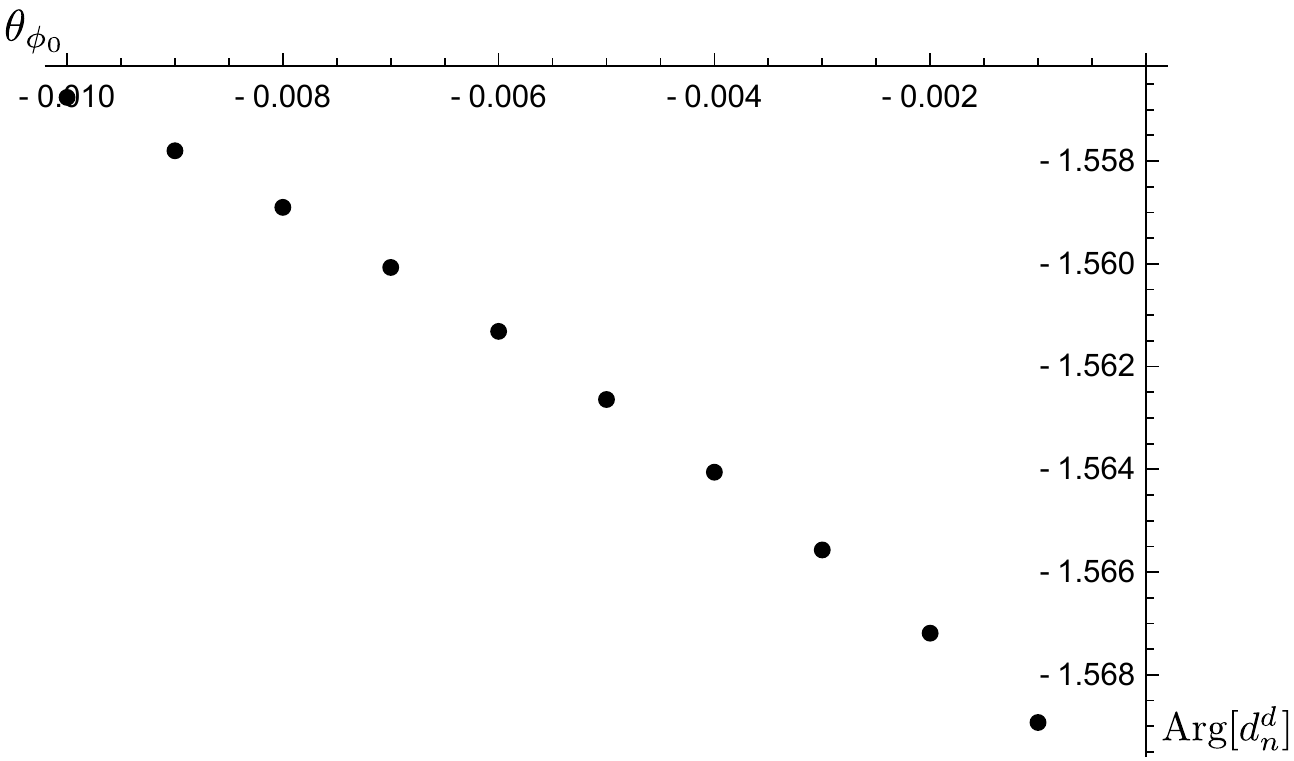}
\end{center}
\caption{Numerical results for the phase of the constant $d^d_n$ as a function of the initial phase of $\phi(0)$, for $m^2 = 0.05$, $n=11$ and $|d_n(0)|=1$, where $|d_n(0)|=1$ is the initial amplitude of the tensor mode. The absolute value of $\phi(0)$ is determined by \eqref{lyonsinitial}.  The plot shows that $\text{Arg}[d^d_n] \simeq - \pi/2 - 1.4 \, \theta_{\phi_0}$.}
\label{tensorphase}
\end{figure}

As before we do the simulation along the contour $\tau = iy$, and deal with the correction due to the curvature of the contour later. 
The results for different values of $n$, $\phi_0$ and $\theta_{\phi_0}$ can be see in figure \ref{tensorsim} and \ref{tensorphase}. They indicate that
\begin{equation}
\frac{d^d_n}{d^g_n} \simeq - i e^{2.1\pi i \epsilon_*}\frac{b^3_*}{3}.
\end{equation}
At first this result might appear wrong since the tensor equation does not involve the scalar field. Nevertheless indirectly, through the background equations of motion, the scalar field interacts with for example the Hubble parameter. Using this result we can analyse the constant part of the tensor action \eqref{tensoraction} in the AdS representation. We find that
\begin{equation}
I_{(n),AdS}^{const} = e^{2.1 \pi i \epsilon_*} \frac{n^3}{2\mathcal{H}_{L*}^2}t^2_{(n)}.
\end{equation}
which is equivalent to the well known dS result \cite{Halliwell} up to slow roll corrections. As for the scalar perturbations the curvature of the contour will give an extra phase to both $d^d_n$ and the derivatives in the action \eqref{tensoraction} of order $\epsilon_*$. In the real part of the action this amounts to corrections of order $\mathcal{O}(\epsilon_*^{-2})$.

\subsection{General potentials}
\label{genpot}
We have seen that the complexity of the no-boundary saddle points induces specific corrections to the perturbation spectra at higher order in the slow roll parameter $\epsilon$. In the limit where $\epsilon$ approaches zero, the AdS domain wall regime of the saddle points becomes real and the corrections vanish. One might wonder whether this conclusion holds more generally and in particular for inflationary potentials with a region of eternal inflation. Such potentials are important because the NBWF predicts that our universe emerges from a region of eternal inflation \cite{Hertog2013}. In eternal inflation the slow roll parameters need not be small. Hence one might expect a significant corrections due to the complexity of the background.

However we will now show that even though this is the case for modes leaving the horizon during eternal inflation, for modes on observable scales the corrections are again small. As an example consider the following hilltop potential  
\begin{equation}
V(\phi) \simeq V_0 - \frac{1}{2} m^2 \phi^2 + \cdots. 
\label{hilltoppot}
\end{equation}
This has a region of eternal inflation near the top of the potential, when $\phi < V_0^{3/2}/m^2$. Slow roll inflation occurs away from this region, in a regime where the higher order corrections in the potential \eqref{hilltoppot} dominate. In this regime the potential \eqref{hilltoppot} obeys the slow roll condition $\epsilon \simeq |V'/V|^2 < 1$. 

\begin{figure}
\begin{center}
\includegraphics[scale=0.7]{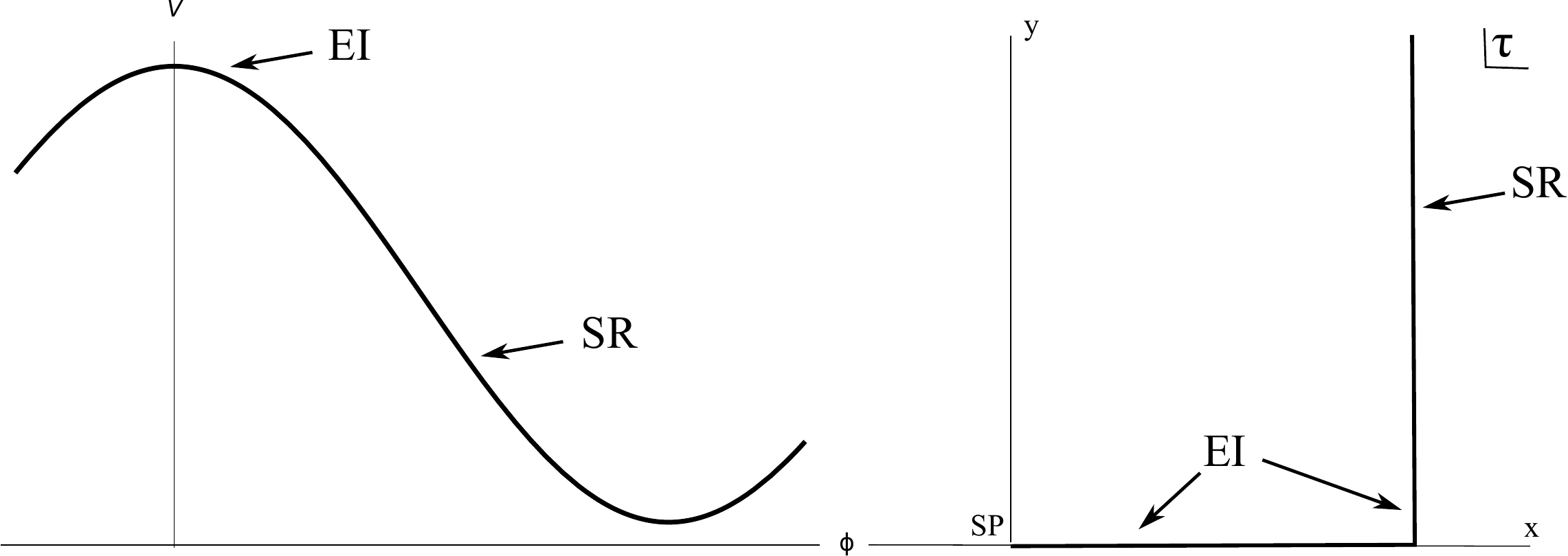}
\end{center}
\caption{Hilltop inflation has both an eternal inflating and a slow roll regime. These regions are indicated in the potential (left) and for the dS contour in the complex $\tau$ plane (right).}
\label{hilltoppic}
\end{figure}

The two different regions are depicted in figure \ref{hilltoppic}. Classical slow roll evolution occurs only high up the vertical part of the dS contour. In this region the field equations \eqref{bgeom1} and \eqref{bgeom2} can be accurately solved up to first order in the slow roll parameter as follows
\begin{align}
a(\tau) &\simeq c_a \exp \left[-i \sqrt{2V(\phi)} \tau \right],\\
\phi(\tau) &\simeq c_\phi + \frac{i V'(\phi)}{3 \sqrt{2V(\phi)}}  \tau. \label{phihill}
\end{align}
The constants $c_a$ and $c_\phi$ are determined by the full evolution from the SP and are thus influenced by the eternal inflation regime. The turning point and the initial phase of the scalar field are chosen such that both $a_{dS}$ and $\phi_{dS}$ are real. The $\phi$ solution \eqref{phihill} is of course not known explicitly, since the potential depends on $\phi$ itself. However, this is not a problem when we only want to estimate the the magnitude of the phase of $\phi$ along the AdS contour, as we will now show. The slow roll regime of the AdS contour lies at
\begin{equation}
x_{adS} = x_{dS} - \frac{\pi}{2\sqrt{2V(\phi_{dS})}}.
\end{equation}
The deviation from a straight contour is directly related to how fast $V$ changes and is therefore small when the slow roll parameter is small. Along this contour the scalar field takes the form
\begin{equation}
\phi_{adS} = \phi_{dS} - \frac{i\pi}{12}\frac{ V'(\phi_{dS})}{ V(\phi_{dS})}.
\end{equation}
Combining this with \eqref{phihill} we can again conclude that the complexity of the AdS domain wall is slow roll suppressed in this region, irrespective of the prior evolution of the fields. Therefore the spectrum of primordial perturbations on observable scales will only receive corrections that are higher order in the small parameter $\epsilon$.

\section{Discussion}
\label{conclusion}

The representation of the no-boundary wave function in terms of saddle point geometries with an approximate AdS interior provides an appealing way to formulate a semiclassical version of dS/CFT using Euclidean AdS/CFT. In this representation the no-boundary measure in cosmology is specified by the regularized action of the interior nearly AdS regime of the saddle points. The matter fields are in general complex in the AdS regime, because otherwise they wouldn't be both real at the final boundary and everywhere regular. The dual form of the no-boundary measure therefore involves the partition function sourced by certain complex deformations of Euclidean CFTs defined on the boundary of the AdS regime. 

The phases of the sources associated with bulk matter fields are tied to the no-boundary condition of regularity and thus encode information about the quantum state of the universe. We have computed the effect of this phase on the predictions for cosmological observables associated with CMB perturbations in backgrounds undergoing an early phase of scalar field driven slow roll inflation. This potentially provides an observational discriminant between a dS/CFT framework based on the NBWF and other dS/CFT proposals. The latter include proposals based on the analytic continuation of real inflationary (asymptotically de Sitter) universes to real Euclidean AdS domain walls which are designed to implement exact Bunch-Davies initial conditions \cite{McFadden2009,McFadden2010,McFadden2010b,McFadden2010c,McFadden2011,Bzowski2011,Bzowski}. In that case the sources in the partition function dual to bulk matter fields are evidently all real.

We have seen that in single field models the predictions for the two-point correlators of scalar and tensor perturbations of both proposals agree to leading order in the slow roll parameters, but they differ at higher order. The agreement at leading order arises because the inflaton is nearly massless during slow roll inflation. This is precisely the special case for which the background matter fields are nearly real in the AdS regime of the NBWF saddle point geometries \cite{Hertog2011}. The different predictions of both proposals at higher order can probably be traced to subtle differences in the quantum state of the fluctuations. The dS/CFT proposals based on analytic continuation are designed such that the fluctuation modes start off in the Bunch Davies vacuum, as in the standard inflationary analysis. By contrast, the initial conditions of the modes in the NBWF are determined by the regularity conditions on the complex saddle points \cite{Hartle2010}. The complex structure of the instanton background means that fluctuations are not exactly in the Bunch-Davies state when the Lorentzian, classical, inflationary evolution emerges. From the analysis in \cite{Hartle2010} it follows that this leads to higher order corrections in the two-point correlators.

It is interesting to ask whether the higher order in $\epsilon$ corrections due to the phase of the background field can potentially constitute an observational signature of the underlying quantum state. In the single field models we discussed in detail these corrections are manifest at order $\epsilon^2$.
At this level they compete with other corrections and would therefore have to be disentangled from those\footnote{However their effect will be much larger than for example loop corrections, which are of order $H^2/M_{pl}^2$ \cite{Woodard2014}.}. The same is true for other observables such as the running of the power spectrum and corrections to higher point functions measuring e.g. non-Gaussianities. In the latter case the third order action $I^{(3)}$ \cite{Maldacena2002}, evaluated along the AdS contour, will similarly acquire corrections of order $\epsilon$ relative to the result based on real backgrounds. Take for example the action for three scalars, which contains a term,
\begin{equation}
I^{3}_{(n)} \sim \int \epsilon^2 a^3 \dot{\zeta}_{(n)}^2 \zeta_{(n)}.
\end{equation}
Each derivative term will both receive a phase due to the perturbation of the AdS contour \eqref{dtau} induced by the phase of the matter and due to the phase of the decaying mode of the perturbations \eqref{cdcgfrac}. The corrections are thus of similar order as for $I^{(2)}$.

It would be interesting to extend our result to the case of multiple fields. It is not implausible that the observational implications due to the phase of the background fields in the NBWF saddle point geometries are more pronounced in multi-field models where the origin of inflation and the perturbations on observable scales are governed by different fields.

\vskip .2in

\noindent{\bf Acknowledgments:} We thank Adam Bzowski, Gabriele Conti, Ruben Monten, Thomas van Riet and Yannick Vreys for discussions. We thank the KITP and the Physics Department at UCSB for their hospitality. This research was supported in part by the National Science Foundation under Grant No. PHY11-25915 and by the National Science Foundation of Belgium (FWO) grant G.001.12 Odysseus. TH is supported by the European Research Council grant no. ERC-2013-CoG 616732 HoloQosmos. We also acknowledge support from the Belgian Federal Science Policy Office through the Inter-University Attraction Pole P7/37 and from the European Science Foundation through the Holograv Network.

\appendix
\section{Perturbation theory}
\label{appendix}
In this Appendix we list some important conventions and equations regarding linear perturbations in a homogeneous and isotropic background. These where first derived for the NBWF in \cite{Halliwell}. The most general metric of scalar and tensor perturbations around the homogeneous isotropic background \eqref{homsad} is
\begin{align}
ds^2 &=(1+2\varphi)d\tau^2 + 2a(\tau) B_{|i}dx^i d\tau + a^2(\tau)[(1-2\psi)\bar{\gamma}_{ij} + 2 E_{|ij} + 2 \gamma_{ij}]dx^i dx^j,
\end{align}
where $\bar{\gamma}_{ij}$ is the metric of the unit radius three-sphere, $x^i$ the coordinates on the three sphere and the vertical bar denotes covariant differentiation with respect to $\bar{\gamma}_{ij}$. Following \cite{Halliwell} we can use the spherical symmetry of the background to expand the perturbations in the normalized scalar harmonics $Q^n_{lm}$ on $S^3$
\begin{align}
\varphi &= \frac{1}{\sqrt{6}} \sum_{nlm} g_{nlm} Q^n_{lm},  &&B = \frac{1}{\sqrt{6}} \sum_{nlm} \frac{k_{nlm} Q^n_{lm}}{(n^2-1)}, \notag \\
\psi &= \frac{-1}{\sqrt{6}} \sum_{nlm} (a_{nlm} + b_{nlm})Q^n_{lm},  &&E = \frac{1}{\sqrt{6}} \sum_{nlm} \frac{3 b_{nlm} Q^n_{lm}}{(n^2-1)}.
\label{metricpert}
\end{align}
In the same manner we can define the perturbation of the scalar field
\begin{equation}
\phi(\tau,x) = \phi(\tau) + \delta \phi(\tau,x),
\end{equation}
where
\begin{equation}
\delta\phi = \frac{1}{\sqrt{6}} \sum_{nlm} f_{nlm} Q^n_{lm}.
\label{fieldpert}
\end{equation}
Finally the tensor perturbation $\gamma_{ij}$ can be expanded in the normalized tensor harmonics $(G_{ij})^n_{lm}$ on $S^3$
\begin{equation}
\gamma_{ij} = \sum_{nlm} d_{nlm}(G_{ij})^n_{lm}.
\end{equation}
We denote the labels n, l and m collectively by $(n)$. 

The equations of motion for the mode functions are
\begin{align}
\ddot{a}_{(n)} &+ 3\mathcal{H} \dot{a}_{(n)} - \frac{1}{3}(n^2-4)a^{-2}(a_{(n)} + b_{(n)}) + 3\dot{\phi}\dot{f}_{(n)} + 3m^2\phi f_{(n)} \notag \\
&\;=  \mathcal{H}  \dot{g}_{(n)} -3 \left(m^2 \phi^2 +1\right) g_{(n)} + \frac{1}{3}(n^2 + 2)a^{-2} g_{(n)} -\frac{1}{3} a^{-1} \dot{k}_{(n)} - \frac{2}{3} \mathcal{H} a^{-1} k_{(n)}, \label{perteom1}\\
\ddot{b}_{(n)} &+ 3\mathcal{H} \dot{b}_{(n)} + \frac{1}{3}(n^2-1)a^{-2}(a_{(n)} + b_{(n)}) \notag \\
&\;= - \frac{1}{3}(n^2-1)a^{-2}g_{(n)}  +\frac{1}{3}a^{-1} \dot{k}_{(n)} + \frac{2}{3} \mathcal{H} a^{-1} k_{(n)}, \\
\ddot{f}_{(n)} &+ 3\mathcal{H} \dot{f}_{(n)} - [m^2 +(n^2-1)a^{-2}]f_{(n)} + 3\dot{\phi}\dot{a}_{(n)} = \dot{\phi}\dot{g_{(n)}} + 2 m^2 \phi g_{(n)} - \phi a^{-1} k_{(n)}, \\
\ddot{d}_{(n)} &+ 3 \mathcal{H} \dot{d}_{(n)} - (n^2-1)a^{-2}d_{(n)} = 0.
\end{align}
The constraint equations are given by
\begin{align}
&\dot{a}_{(n)} + \frac{(n^2-1)}{(n^2-4)}\dot{b}_{(n)} + 3 \dot{\phi} f_{(n)} = \mathcal{H} g_{(n)} - \frac{1}{(n^2-1)}a^{-1} k_{(n)}, \label{perteom4} \\
&3a_{(n)}(\dot{\phi}^2 -\mathcal{H}^2) + 2(\dot{\phi} \dot{f}_{(n)} - \mathcal{H} \dot{a}_{(n)}) -m^2(2\phi f_{(n)} + 3 \phi^2 a_{(n)}) - 3a_{(n)} \qquad &\nonumber \\
& \qquad + \frac{2}{3}a^{-2}[(n^2-4)b_{(n)} +(n^2+1/2)a_{(n)}] = \frac{2}{3} \mathcal{H} a^{-1} k_{(n)} + 2 g_{(n)} \left(\dot{\phi}^2 - \mathcal{H}^2 \right). \label{perteom5}
\end{align}
The above scalar degrees of freedom are not gauge invariant. There are two gauge degrees of freedom characterized by the Lagrange multipliers $g_{(n)}$ and $k_{(n)}$, which allows us put two degrees of freedom to zero. An often used gauge is the Newtonian gauge $b_{(n)} = k_{(n)} =0$. The remaining three variables are related to each other trough the linear Hamiltonian and momentum constraints \eqref{perteom4} and \eqref{perteom5}. The physical perturbation corresponds to a single gauge invariant variable $\zeta$, the curvature perturbation on comoving hyper-surfaces, given by the linear combination \cite{Bardeen1980}
\begin{equation}
\zeta_{(n)} = a_{(n)} + b_{(n)} - \frac{\mathcal{H}}{\dot{\phi}} f_{(n)}.
\end{equation}
The equation of motion for $\zeta_{(n)}$ can be derived from the above, when taking the gauge $b_{(n)} = f_{(n)} = 0$. In this case $\zeta_{(n)} = a_{(n)}$, whose equation of motion can be obtained by eliminating $g_{(n)}$ and $k_{(n)}$ using the constraint equations. The result is
\begin{align}
&\left[3\dot{\phi}^2 + \mathcal{H}^2(n^2-4) \right] \ddot{\zeta}_{(n)}  + \left[9\dot{\phi}^2 +(n^2-4)\left(3\mathcal{H}^2 +6\dot{\phi}^2 + 2a^{-2} +2\mathcal{H} \frac{\ddot{\phi}}{\dot{\phi}} \right)\right]\mathcal{H} \dot{\zeta}_{(n)} \nonumber \\
&\;\; + \left[3\dot{\phi}^2 +(n^2-4)\left(\mathcal{H}^2 +3\dot{\phi}^2 + 2a^{-2} +2\mathcal{H} \frac{\ddot{\phi}}{\dot{\phi}}  -\mathcal{H}^2(n^2-4)\right)\right]a^{-2} \zeta_{(n)} =0.
\label{zetaeom}
\end{align}
The action for a scalar mode $\zeta_{(n)}$ was derived by \cite{Shirai} and is just a boundary term
\begin{equation}
I_{(n)} = a M \bar{z}_{(n)} \dot{\bar{z}}_{(n)} - N \bar{z}_{(n)}^2,
\end{equation}
where $\bar{z}_{(n)} \equiv a^3 \dot{\phi} z_{(n)}$, the dot denotes a derivative with respect to $\tau$, perpendicular to the boundary, and
\begin{align}
M &\equiv \frac{(n^2-4)}{2a^4[(n^2-4)\mathcal{H}^2 + 3 \dot{\phi}^2]}, \\
N &\equiv \frac{1}{4 M U a^3} \left[K \left(2a^4 - 3a^6m^2\phi^2 +3 \frac{n^2-1}{n^2-4}a^6\dot{\phi}^2 \right) + a^{12} m^4 \phi^2 +3 a^{12} \mathcal{H} \phi \dot{\phi} \right], \label{N}\\
U &\equiv K a^3 \mathcal{H} + a^9 m^2 \phi \dot{\phi}, \label{U}\\
K &\equiv \frac{1}{3}\left[(n^2-4)a^4 \mathcal{H}^2 -(n^2+5)a^6 \dot{\phi}^2 -(n^2-4)a^6 m^2 \phi^2 \right], \label{Kn}
\end{align}
where all the quantities are to be evaluated at the boundary. This action in valid in the regime where the cosmological constant term is negligible. To re-implement the influence of the cosmological constant, one can make the substitution $m^2 \phi^2 \rightarrow m^2 \phi^2 + H^2$.

The action for a tensor mode $t_{(n)}$ is also a boundary term
\begin{equation}
I_{(n)}^{(2)} = \frac{1}{2}a^3 t_{(n)} \dot{t}_{(n)} + 2 a^3 \mathcal{H} t_{(n)}^2,
\end{equation}
where again all the quantities are to be evaluated at the boundary.

\bibliographystyle{jhep}
\bibliography{references1}

\end{document}